\documentclass[aps,pre,citeautoscript,singlecolumn,showkeys, floatfix, superscriptaddress]{revtex4-1}
\pdfoutput=1
\usepackage{amsmath}
\usepackage{hyperref}
\usepackage{mathtools}
\usepackage{empheq}
\usepackage{graphicx}
\usepackage{float}
\usepackage{natbib}
\usepackage{epstopdf}
\usepackage{amsthm}
\usepackage{xcolor}
\usepackage{textcase}
\usepackage[normalem]{ulem}

\newcommand*{\myfont}{\fontfamily{lmr}\selectfont}

\font\msytw=msbm9 scaled\magstep1

\let\a=\alpha   
   
 \let\m=\mu   
\let\s=\sigma

   \let\io=\infty

\def\RRR{{\cal R}}  
 \def\SS{{\cal S}}

\def\to{\rightarrow}  

\def\RRR{\hbox{\msytw R}}

\newcommand{\beq}{\begin{equation}} \newcommand{\eeq}{\end{equation}}

\begin{document}

\title{Self-planting: digging holes in rough landscapes}

\author{Dhruv Sharma}
\affiliation{Laboratoire de Physique de l'Ecole Normale Sup\'erieure, ENS, Universit\'e PSL, CNRS, Sorbonne Universit\'e, Universit\'e de Paris, Paris, France}

\author{Jean-Philippe Bouchaud}
\affiliation{
CFM, 23 rue de l'Universit\'e, 75007 Paris, France
}


\author{Marco Tarzia}
\affiliation{
	LPTMC, CNRS-UMR 7600, Sorbonne Universit\'e, 4 Pl. Jussieu, F-75005, Paris, France
}
\author{Francesco Zamponi}
\affiliation{Laboratoire de Physique de l'Ecole Normale Sup\'erieure, ENS, Universit\'e PSL, CNRS, Sorbonne Universit\'e, Universit\'e de Paris, Paris, France}

\begin{abstract}
Motivated by a potential application in economics, we investigate a simple dynamical scheme to produce planted solutions in optimization problems with continuous variables. We consider the perceptron model as a prototypical model. Starting from random input patterns and perceptron weights, 
we find a locally optimal assignment of weights by gradient descent; we then remove misclassified
patterns (if any), and replace them by new, randomly extracted patterns. This ``remove and replace'' procedure is iterated until perfect classification is achieved.
We call this procedure ``self-planting'' because the ``planted'' state is not pre-assigned but results from a co-evolution of weights and patterns.
We find an algorithmic phase transition separating a region in which self-planting is efficiently achieved from a region in which it takes exponential time in the system
size. We conjecture that this transition might exist in a broad class of similar problems. 
\end{abstract}

\maketitle

\section{Introduction}
\label{sec:intro}


\paragraph{The perceptron --}
The perceptron was one of the first formal models of how neurons function. Starting with the pioneering work of Rosenblatt~\cite{Rosenblatt1958}, it has continued to be both a paradigm and a cornerstone of machine learning. Statistical physicists became interested in the perceptron problem and significant progress was made in the late 1980's in a series of papers~\cite{Gardner1988,Gardner1988a,Brunel1992,Krauth1988,Krauth1989}. Recent interest in the perceptron surged again after Franz and Parisi proposed it as a toy model to understand the jamming of hard spheres~\cite{Franz2016}. The perceptron can indeed be seen both as a generic continuous constraint satisfaction problem or as a learning/classification problem. 
To be precise, one considers the following constraint satisfaction problem for vectors $\vec{X}$ on the $N$-dimensional sphere $\mathcal{S}_{N}$, such that $|\vec X|^2 = N$. 
Given $M$ vectors $\vec{\xi}_{\mu} \in \RRR^N$, 
we look for $\vec{X} \in \SS_N$ such that:
\begin{align}\label{define_h_mu}
h_{\mu}(\vec{X}):=\frac{1}{\sqrt{N}} \vec \xi_{\mu} \cdot \vec X -  \sigma > 0 \qquad \forall \mu \in \left \{ 1 \ldots M \right\} \, .
\end{align}
The quantities $h_{\mu}$ will be called gaps in the following.

Several problems can be encoded by this set of constraints.
\begin{itemize}
\item
 For positive $\sigma$, this corresponds to the standard perceptron problem (more precisely, to a support vector machine)~\cite{Friedman2001} for classification. 
 In fact, consider input vectors $\vec{\Xi}_\m\in\RRR^N$ and corresponding outputs $y_\m = f(\vec\Xi_\m)\in\{\pm 1\}$. If all the constraints defined by Eq.~\eqref{define_h_mu} with $\vec\xi_\m = y_\m \vec\Xi_\m$ are satisfied,
 then there exists a hyperplane $\vec\Xi \cdot \vec W=0$ (with $\vec W = \vec X/\sqrt{N}$ and $|\vec W|^2=1$) which correctly classifies all inputs according to the rule $y_\m = \text{sgn}(\vec W \cdot \vec \Xi_\m)$, and moreover the minimal distance $d_\m=y_\m \vec\Xi_\m \cdot \vec W$ of points $\vec\Xi_\m$ from that hyperplane
 is greater than $\s$ (hence, $\s$ quantifies the robustness of the classification).

 \item 
One can ask a dual question:  given $M$ random pairings of inputs and outputs ($\vec{\Xi}_{\mu}$, $y_{\mu}$), what is the maximal number of such pairings that a perceptron can learn? Or formulated differently, in the absence of a relation $f(\vec{\Xi}_{\mu})$ relating $y_{\mu}$ and $\vec{\Xi}_{\mu}$, how many random patterns can the perceptron classify? This is known as the storage problem and the critical number $M_{c}$ of patterns learnt is noted as the storage capacity ~\cite{Gardner1988a,Gardner1988}.
\item
  When $\sigma < 0$, the problem can no longer be interpreted as a classification problem, but it is still a non-convex constraint satisfaction problem. 
  In particular, it provides an extreme schematisation of the sphere jamming problem, in which one substitutes the interaction between spheres with a random background of spherical obstacles placed at random positions in space~\cite{Franz2016}.
  As for any constraint satisfaction problem, one can transform it to an optimization problem by introducing a cost function that assigns a positive cost (or energy) to a violated constraint.
  With random $\vec\xi_\m$, this latter optimization problem has been studied in great detail and a rich phase diagram has been computed \cite{Franz2017a}. Briefly, the phase diagram has a SAT phase, in which all the constraints in Eq.~\eqref{define_h_mu} are satisfied, and an UNSAT phase, in which at least one of the $h_{\mu}$ is unsatisfied, as illustrated in Fig.~\ref{fig:phase_diagram}. Each of these phases is further sub-divided into regions where replica symmetry is maintained (RS), suggesting a simple energy landscape characterized by a unique global minimum, and regions where replica symmetry is broken (1-RSB or full-RSB), suggesting a rough energy landscape full of local minima.
\end{itemize}

\paragraph{Planting, quiet planting, and the teacher-student scenario --} 
 In the context of constraint satisfaction problems, it is often interesting to consider problems that are SAT by construction, i.e. admitting at least one solution. This can be achieved by the simple ``planting'' technique~\cite{flaxman2008random}.
 In the case of the perceptron problem, 
  it consists in choosing a random state
 $\vec{X}^{\text{plant}}$ and then independently extracting patterns $\vec\xi_\m$ such that the constraint $h_\m(\vec{X}^{\text{plant}}) > 0$ is satisfied. Because each constraint is considered independently, the construction of such planted instances is computationally easy. One can for instance
 extract $\vec\xi_\m$ from the normal distribution $\mathcal{N}(0,1)$ and, if the constraint is violated, reject it and extract a new one, until the constraint is satisfied. Because the probability of satisfying a constraint is finite and equal to $\text{erfc}(\sigma/\sqrt{2})/2$, the process converges after a few steps.

Note that for the perceptron problem with $\sigma=0$, the planting procedure also essentially coincides with the so-called ``teacher-student'' learning scenario~\cite{Engel2001} (also known as the ``generalization problem''). One can define a teacher $\vec T \in \RRR^N$ with $|\vec T|^2=N$, extract random i.i.d. normal patterns $\vec\Xi_\m$,
and define outputs $y_\m = \text{sign}(\vec T \cdot \vec\Xi_\m)$. The perceptron problem with patterns $\vec\xi_\m = y_\m \vec\Xi_\m$ then necessarily admits a solution $\vec{X}^{\text{plant}} = \vec T$.
 
 One interesting question is whether the ensemble of such random planted problems is distinguishable from the completely random ensemble in which the $\vec\xi_\m$ are i.i.d. normal variables. In the context of learning, these two ensembles occur as specific cases of a more general teacher-student scenario in which the teacher (or reference) perceptron $\vec{T}$ is itself noisy~\cite{Gyorgyi1990a}. Tuning the level of noise of the teacher perceptron then allows one to interpolate between the purely deterministic case, where $\vec{T}$ is exactly known, and the case of infinite noise, which corresponds to the storage capacity problem. 

 The random planted ensemble and the purely random ensemble have very different properties: the disorder $\vec{\xi}_{\mu}$ in the planted ensemble is in fact correlated since it depends on the planted configuration $\vec{X}^{\text{plant}}$ ~\cite{Zdeborova2018}. Nevertheless it has been shown that in certain regions of parameter space,
 the typical realizations of the patterns in the two ensembles have statistically identical properties, and the planted ensemble thus coincides with the random one. This construction, called ``quiet planting''~\cite{Barthel2002,Krzakala2009,Krzakala2014,Zdeborova2018}, 
 is particularly powerful (e.g. as an algorithmic benchmark)
 because one can construct
instances of constraint satisfaction problems for which a solution is known, but are otherwise representative of the random ensemble.

\paragraph{Self-planting --}

Optimization problems are generally studied with a fixed free-energy landscape, fixed either by a non-convex loss function or via quenched disorder. In this work, we extend our enquiry to dynamically changing landscapes. We ask how the structure of the minima of an optimization problem evolve as the energy landscape itself is modified by an external process. This leads to dynamic, non-quenched disorder. In this work we study the particular case of the perceptron where the $\xi_{\mu}^{i}$ are dynamically updated, with a rule that seeks to improve violated constraints. We call the procedure ``Remove \& Replace'' (R\&R in the following), as a generalization of the planting procedure -- see section \ref{sec:rr_dynamics} for a more precise statement. 

In this present paper, we detail how the the  R\&R dynamical rule allows the system to reach a SAT configuration through the co-evolution of both $\vec X$ and the $\vec{\xi}_{\mu}$, only when $\sigma < \sigma_{\rm RR}$, where $\sigma_{\rm RR}$ is an algorithmic threshold. This scenario is similar to planting, but in our case, the planted solution 
$\vec{X}$ is not chosen {\it a priori} but is the result of our co-evolution rule -- hence our proposal to call this effect ``self-planting''. We also highlight how the landscape evolves as we cross the $\sigma_{\rm RR}$ threshold and compare the situation with that of the static case, i.e. quenched disorder.

We believe that a similar phenomenon should hold in different situations of interest (in which the connectivity of the constraints is large or infinite), 
such as hard spheres in large dimensions~\cite{charbonneau2017glass} or the Hopfield model~\cite{Hopfield1982}, and, as discussed in the conclusion, for non-classifiable data.

\paragraph{The perceptron in an economic context --}

In a forthcoming paper ~\cite{Sharmaa}, we shall present our interpretation of the perceptron as a stylized model for the economy. For completeness, we cursorily describe the model here. Consider $M$ agents and $N$ products with prices $\vec{p} \in \RRR^N$. Each agent, labeled by $\mu \in \left \{ 1 \ldots M \right\} $, wants to buy or sell a quantity $\xi_{i}^{\mu}$ (positive for buy and negative for sell) of product $i$, with the constraint $\sum_{\mu=1}^{M} \xi_{i}^{\mu}= 0$, $\forall i$, which ensures market clearing. The quantity $\pi_{\mu} = \sum_{i=1}^{N} \xi_{i}^{\mu} p_{i}$ is then the total money spent or earned by agent $\mu$. For each agent $\mu$ we impose the constraint $\pi_{\mu} \geq b_{\mu}$. Here, $b_{\mu} >0$ if the agent is only allowed to make profits, or $b_{\mu} < 0$ if the agent is allowed to borrow. This budgetary constraint is then equivalent to the perceptron problem \ref{define_h_mu}. 

In an economic context, the R\&R procedure corresponds to the removal of agents who are bankrupt (i.e. have violated their constraints) and their subsequent replacement with other agents (or maybe the same agents but with different ``preferences'' $\xi_{i}^{\mu}$).

\section{The Model and its Phase Diagram}
\label{sec:model}

In order to study the model in the UNSAT phase, we introduce the Hamiltonian
\beq
H_{0}(\vec{X}) = \frac{1}{2} \sum_{\mu=1}^{M} h_{\mu}^2 \theta(-h_{\mu}), 
\label{define_hamiltonian}
\eeq
which can be thought of as a cost function penalizing violated constraints in Eq.~\eqref{define_h_mu}. The SAT phase then corresponds to the situation where the ground-state energy is exactly zero. When the minimum of $H$ is strictly positive, there is at least one unsatisfied constraint for any choice of $\vec{X}$ and we are in the UNSAT phase. From the statistical physics point of view, the zero-temperature phase diagram of the model (i.e. the structure of minima of the Hamiltonian) has been well studied in a series of works~\cite{Franz2016,Franz2017a}. 

\begin{figure}[t]
\includegraphics[width=.6\textwidth]{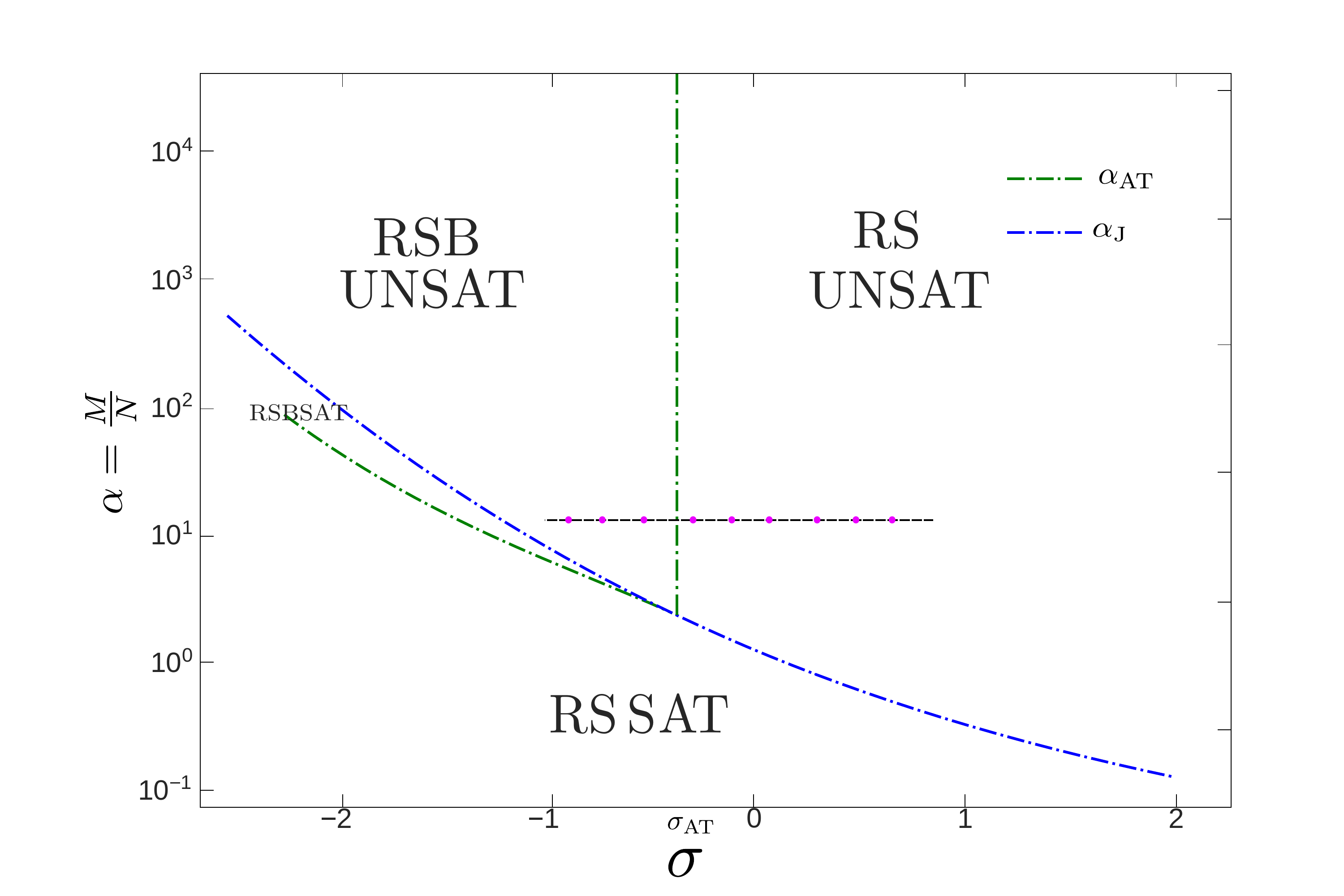}
\caption{Thermodynamic zero-temperature phase diagram for the perceptron problem with $m=0.577$. 
$\alpha_{\text{J}}$ is the SAT-UNSAT transition line, $\alpha_{\text{AT}}$ is the de Almeida-Thouless line. Note that the AT line in the UNSAT phase is shifted with respect to $m=0$, and is located at a value of $\sigma_{\rm AT} < 0$. 
There is also a region of the phase diagram at lower values of $\sigma$ (outside the range of the figure) where the dAT line coincides with a continuous transition from a replica symmetric SAT phase
to a stable 1RSB SAT phase.  At even lower values of $\sigma$ a RFOT type phenomenology is observed, with a dynamical and a Kauzmann transition to a 1RSB SAT phase ~\cite{Franz2017a}.
Note that the SAT-UNSAT threshold is computed within the replica symmetric approximation, and hence is exact only for
$\sigma >\sigma_{\rm AT}$.
The dotted black line corresponds to a cut at constant $\alpha=20$: keeping $\alpha$ fixed we perform R\&R dynamics at various values of $\sigma$.}
\label{fig:phase_diagram}
\end{figure}

Compared to the standard perceptron specification, our economics interpretation leads to a slight variation: apart from the spherical constraint on the $X_{i}$, we will also impose that the sum of the components of each of the vectors be positive: $\frac{1}{N}\sum_{i} X_{i} = m$, i.e., that the magnetisation (or the average price, within the economic interpretation) is positive. The Hamiltonian is hence modified to:
\begin{align}
H_{1}(\vec{X}) &= \frac{1}{2} \sum_{\mu=1}^{M} h_{\mu}^{2} \theta(-h_{\mu}) - \Upsilon \sum_{i=1}^{N} X_{i}, 
\label{modified_hamiltonian}
\end{align}
where $\Upsilon$ acts as a magnetic field. The field $\Upsilon$ and the magnetization $m$ are related by a simple Legendre transform. One can then solve the model in the thermodynamic limit - $M,N \to \infty$ with $\alpha = \frac{M}{N}$ fixed - and retrieve the phase diagram similar to the standard perceptron. The addition of a magnetic field does not change the qualitative features of the phase diagram, which is illustrated in Fig.~\ref{fig:phase_diagram} for $m=0.577$ and can be compared with the one for $m=0$ reported in Ref.~\cite{Franz2017a}. 
The SAT-UNSAT transition line separates the SAT and UNSAT regions, and 
each of these regions is separated in a replica symmetric (RS) region and a replica symmetry broken (RSB) one. 
\begin{itemize}
\item In the RS-UNSAT phase, instances are typically not satisfiable, and the Hamiltonian has a unique minimum at positive energy.
\item In the RSB-UNSAT phase, instances are again unsatisfiable, but the energy landscape is now rough, displaying many local minima, all at positive energy.
\item In the RS-SAT phase, instances are satisfiable, and the space of solutions is simply connected (i.e. the energy landscape has a flat connected bottom at zero energy).
\item In the RSB-SAT phase, instances are satisfiable, but the space of solutions is disconnected in many ``clusters'' of solutions.
\end{itemize}

The de Almeida-Thouless (AT) line ~\cite{Almeida1978} $\sigma_{\text{AT}}(\a)$ 
which determines the RS-RSB boundary is independent of $\a$ (i.e., it is vertical in Fig.~\ref{fig:phase_diagram}) in the UNSAT phase,
and the corresponding value of $\sigma_{\text{AT}}$ is the solution of
\begin{align}
1-m^2 &= \frac{ \int_{-\infty}^{\sigma_{\text{AT}}}(h-\sigma_{\text{AT}})^2 e^{-h^2/2} \mathrm{d}h}{\int_{-\infty}^{\sigma_{\text{AT}}} e^{-h^2/2} \mathrm{d}h}.
\label{compute_sigma_star}
\end{align}
Note that $\sigma_{\text{AT}}=0$ for $m=0$ but $\sigma_{\text{AT}}< 0$ for $m > 0$. The RSB-UNSAT phase corresponds to 
$\sigma < \sigma_{\text{AT}}$.

\section{``Remove and Replace'' Dynamics}
\label{sec:rr_dynamics}

The model is fully specified by the choice of parameters $m$, $\sigma$ and $\alpha$ and the value of $N$. The dynamical process that we propose is then the following:
{\myfont
\begin{enumerate}
\item Start with randomly chosen vectors, $\vec{\xi}_{\mu}$ from the normal distribution $\mathcal{N}(0,1)$, and $\vec{X}$ uniformly on the sphere $|\vec X|^2= N$.

\item Find $\vec{X}^{\text{opt}}$ which (locally) minimizes the cost function $\mathcal{L}$, using a gradient descent algorithm, where $\mathcal{L}$ is defined as 
 \begin{equation}
 \label{cost_function}
 \mathcal{L} = H_{0}(\vec{X}) + \frac{\lambda_1}{2}\Big( \frac{1}{N} \sum_{i=1}^{N} X_{i}^2 - 1 \Big) + \frac{\lambda_{2}}{2} \Big( \frac{1}{N} \sum_{i=1}^{N} X_{i} - m \Big)^2 .
 \end{equation}
Here $\lambda_{1}$ and $\lambda_{2}$ are Lagrange multipliers which fix the spherical and linear constraints that $\vec{X}$ must satisfy.

\item \textbf{Remove} those vectors $\vec{\xi}_{\mu}$ such that $h_{\mu}(\vec{X}^{\text{opt}}) < 0$. These are the constraints that are unsatisfied.

\item \textbf{Replace} the removed vectors $\vec{\xi}_{\mu}$ with new random vectors sampled from the normal distribution $\mathcal{N}(0,1)$.

\item Repeat steps 2-4. Stop if all the constraints are satisfied at step 3.
\end{enumerate}
}
Because this dynamical scheme involves the removal and replacement of the vectors $\vec{\xi}_\mu$, we call this procedure Remove \& Replace. Albeit quite simple, 
this dynamical scheme uncovers a rich picture which we elucidate below. 
Note that if $\vec X$ were not updated in step 2, this procedure would coincide with the standard planting procedure -- so in that sense R\&R should rather be called ``Optimize, Remove \& Replace''.

In the following, we will simulate the R\&R dynamics and focus on the following dynamical observables, which are all functions of the unsatisfied constraints or the ``negative'' gaps. These are the (intensive) energy,
\beq
e(t) = \frac{H_{0}(\vec{X}(t))}N = \frac{1}{2N} \sum_{\mu=1}^{M} h_{\mu}(t)^2 \theta(-h_{\mu}(t)) \ ,
\eeq
the average gap,
\beq
h(t) = - \frac{1}{M} \sum_{\mu=1}^{M} h_{\mu}(t) \theta(-h_{\mu}(t)) \ ,
\eeq
and the average number of ``contacts'',
\beq
z(t) =  \frac{1}{M} \sum_{\mu=1}^{M}  \theta(-h_{\mu}(t)) \ .
\eeq
Note that we define time $t$ as the number of the R\&R steps that have been performed. Yet, the reader should keep in mind that a single R\&R step requires
a number of computer operations that scales as ${\cal C} \times N \times M$, with a large constant ${\cal C}$,
mostly due to the gradient descent. In fact, the evaluation of the gradient of the Hamiltonian takes $N \times M$ operations: one must evaluate the $M$ quantities
$h_\mu$, each of which is a sum of $N$ terms; and this evaluation must be repeated a large number of times for the gradient descent to converge to the local optimum.

\begin{figure}
	\includegraphics[width=0.7\textwidth]{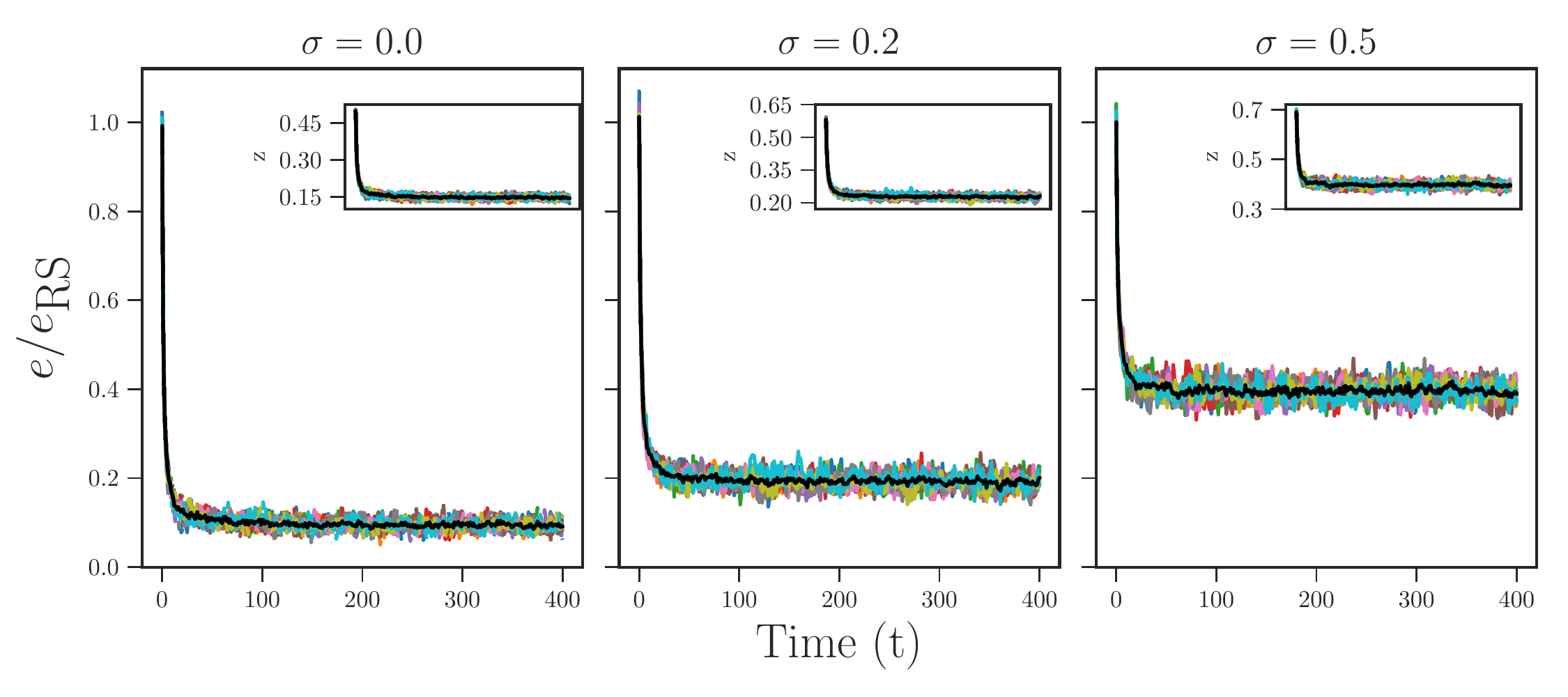}
	\includegraphics[width=0.7\textwidth]{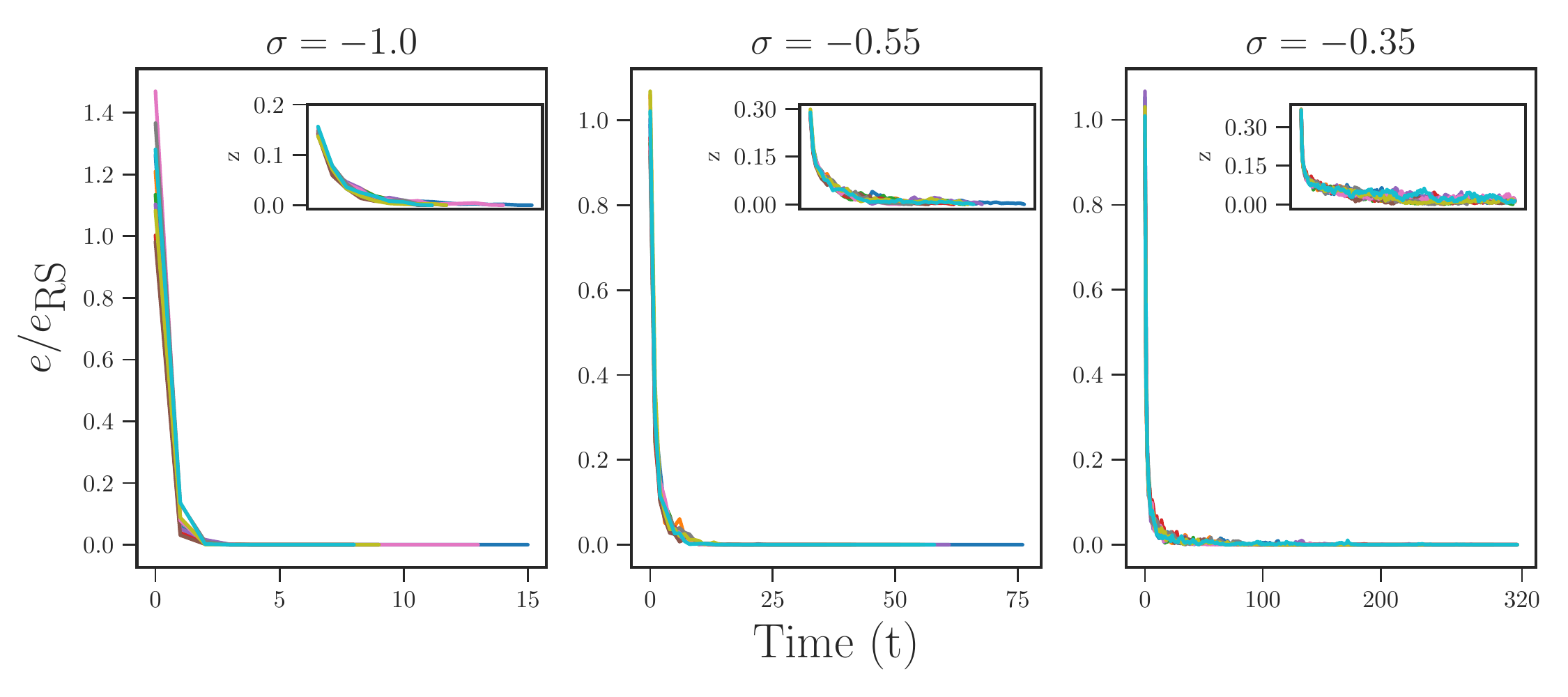}
\caption{Variation of energy $e$ as a function of time $t$ in R\&R dynamics, for $\a=20$, $m=0.577$, $N=200$, and several values of $\s$. 
Individual curves for 10 runs are shown, together with their average (full black line). The inset shows the evolution of the fraction of unsatisfied
constraints, $z$, as a function of time $t$. Top panel: R\&R-UNSAT phase. Bottom panel: R\&R-SAT phase.}
\label{fig:energy_vs_time}
\end{figure}

\section{Results}
\label{sec:org862b6cc}

In this section, we present numerical simulation results for the R\&R dynamics. 
The control parameters are $\a,\s,m,N$ and 
unless otherwise noted, we fix $\a=20$ and $m=0.577$.
Note that if we start at low enough $\s$, all the constraints are satisfied with very high probability (going to one exponentially in $N$)
already in the initial state, and the R\&R dynamics stops before any removal is made. Hence, we only consider large values of $\s$ belonging to
the thermodynamically UNSAT phase in the phase diagram of Fig.~\ref{fig:phase_diagram}, in which instances are guaranteed to be UNSAT with probability
one for $N\to\io$.

\subsection{Long-time behavior of the energy}

As a first step we discuss the long-time behavior of the system under the R\&R dynamics. Fig.~\ref{fig:energy_vs_time} 
shows the energy $e = H/N$ as a function of time for several values of $\s$. 
We rescale the energy by the thermodynamic value $e_{\text{RS}} > 0$, which corresponds to a random choice of the $\vec{\xi}_\mu$. In the RS phase this corresponds to the value
of the energy at time $t=0$. In the RSB phase, the value of $e_{\text{RS}}$ is not accurate, but still provides a decent approximation
to the thermodynamic energy.
Note that in the initial RS state, there is a unique configuration $\vec{X}^{\text{opt}}$, which minimizes the energy for a given set of vectors $\vec{\xi}_\mu$,
while in the RSB phase there are many local minima that trap the gradient descent. This should remain true at least in the initial stages of the R\&R dynamics, in which the energy landscape is not yet altered with respct to $t=0$.
We discuss the long-time evolution of the energy landscape under the R\&R dynamics in a later section.

As the vectors $\vec{\xi}_\mu$ are removed and replaced, we observe that both $e/e_{\text{RS}}$ and the fraction of unsatisfied constraints $z$ are reduced, as expected. Interestingly, for large enough $\sigma$, R\&R is unable to find a configuration such that the constraints are satisfied. After a brief transient, the energy fluctuates around some plateau value which we note $e^\infty$. This long-time value $e^\infty/e_{\text{RS}}$ decreases upon decreasing $\sigma$ as shown in Fig.~\ref{fig:energy_vs_time}, top panel. For example, when $\sigma=0$ the R\&R algorithm finds configurations that are ten times ``better'' than random (i.e. $e^\infty/e_{\text{RS}} \approx 0.1$).


Upon further decreasing $\sigma$, we reach a region in which $e^\infty$ is strictly zero -- see Fig.~\ref{fig:energy_vs_time}, bottom panel. 
The presence of a zero-energy final state implies that we have been able to generate a satisfiable instance of the problem starting from an unsatisfiable one. Furthermore, this zero-energy state is often (see discussion below) reached fairly quickly, within a few steps of the R\&R dynamics, at least when $\sigma$ is sufficiently small.
We denote the final state as $\vec{X}^\infty$ and the corresponding configuration of disorder as $\vec{\xi}_\mu^\infty$. The pair ($\vec{\xi}_\mu^\infty, \vec{X}^\infty$) is a \emph{self-planted} state. 

\subsection{Finite size scaling and R\&R transition}

We expect that a sharp {\it algorithmic} phase transition separates two phases in the limit $N\to\io$: a R\&R-UNSAT phase at large $\sigma$, in which the R\&R dynamics is unable to find a self-planted state, and a R\&R-SAT phase at low $\sigma$, in which the R\&R dynamics converges to a self-planted state in a finite time.
More precisely, we expect to find, for finite but large $N$, 
a scenario similar to the one investigated in ~\cite{Semerjian2003} for the WalkSAT algorithm for the standard Boolean $K$-satisfiability problem:
\begin{itemize}
\item
In the R\&R-UNSAT phase, the dynamics would typically converge to a stationary state with positive energy for $N\to\io$. For finite $N$, fluctuations
around the typical state are controlled by a large deviation function, $P(e) \sim \exp[-N \omega(e)]$. The function $\omega(e)$ attains a maximum at the typical value of $e$, and is finite for $e=0$; therefore, there is an exponentially small
probability in $N$ to reach zero energy by a fluctuation which produces a rare, specific SAT realization of the constraints.
Because $e=0$ is an absorbing state of the dynamics, the R\&R procedure would stop at that point. A self-planted state is then found even
in the R\&R-UNSAT, provided one waits a time $t \sim \exp[N \omega(e=0)]$ growing exponentially with $N$.
\item
In the R\&R-SAT phase, on the contrary, the R\&R dynamics typically reaches a self-planted state in a finite time, even when $N\to\io$. In this case
the function $\omega(e)$ is maximal at $e=0$.
\end{itemize}
One could then hope to identify the transition by looking at the scaling with $N$ of the time $T_{\text{sp}}$ needed to reach a self-planted state; this time
should be finite for $\sigma<\sigma_{\rm RR}$ and diverge exponentially in $N$ for $\sigma>\sigma_{\rm RR}$. Unfortunately, because the perceptron
is a fully-connected model, numerical simulation of the model scales as $N^2$ (to be compared with standard dilute models such as the one of
Ref.~\cite{Semerjian2003}, for which simulations scale as $N$), which limits us to the study of small systems, $N \leq 256$.

\begin{figure}
	\includegraphics[width=0.49\textwidth]{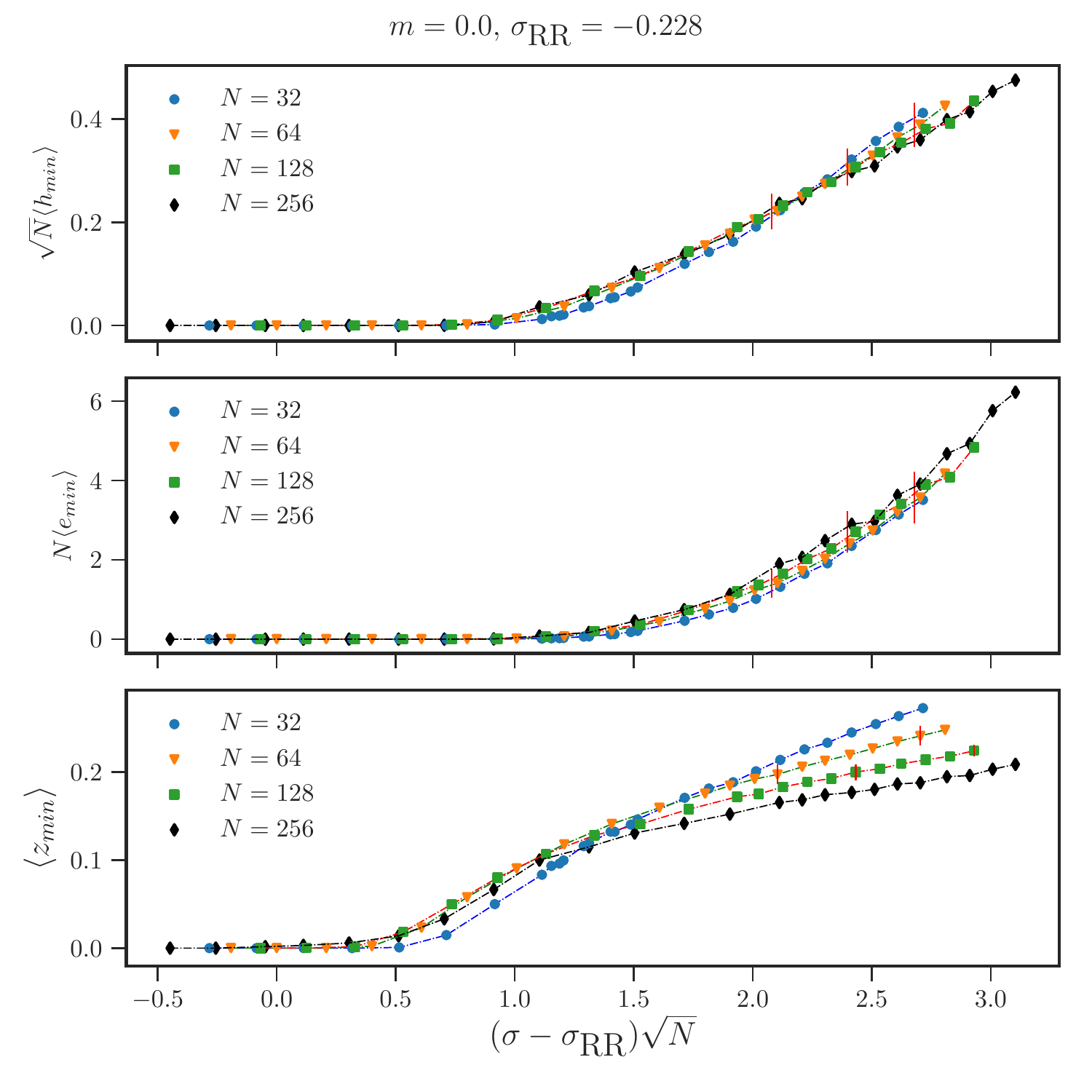}
	\includegraphics[width=0.49\textwidth]{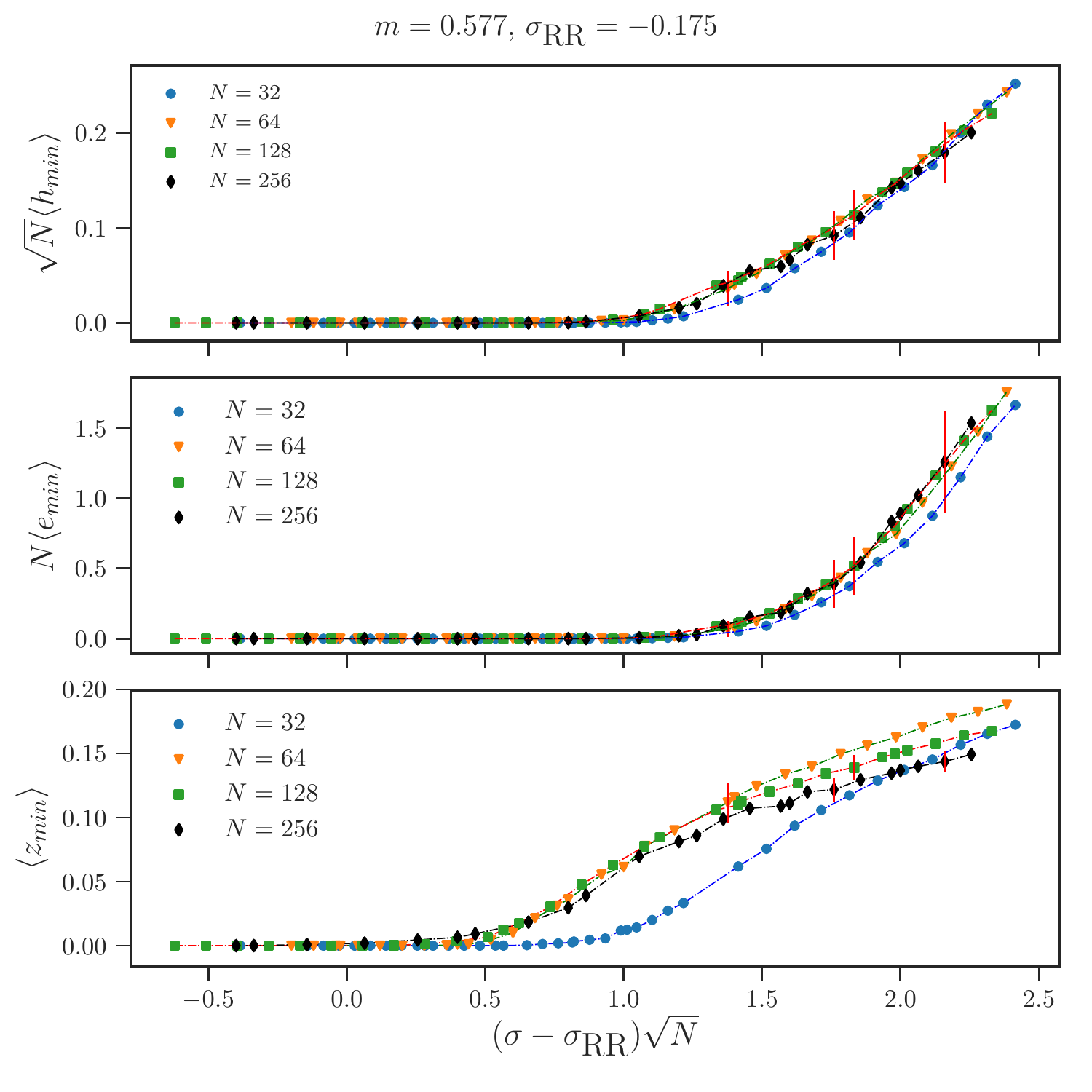}
\caption{
Finite size scaling around the algorithmic transition. For fixed $\a=20$ and $m=0$ (left) or $m=0.577$ (right), we plot the minimal
energy, gap and fraction of contacts observed during the first $T_{\text{max}}=600$ iterations of R\&R, averaged over several realizations.
Data for different $N$ are scaled to achieve collapse by a single parameter $\sigma_{\rm RR}$. For better visualization, error bars are shown only
for a few data points. Apart from the data for $N=32$, the quality of the rescaling is acceptable.
}
\label{fig:FSS1}
\end{figure}

Yet we found that a finite size scaling analysis is possible even for small sizes, $N \gtrsim 64$. We consider R\&R runs of fixed time $T_{\text{max}}=600$ and we compute the minimal
energy $e_{\text{min}} = \min_{t\leq T_{\text{max}}} e(t)$, the minimal average gap $h_{\text{min}} \min_{t\leq T_{\text{max}}} h(t)$, and the minimal fraction of
contacts  $z_{\text{min}}= \min_{t\leq T_{\text{max}}} z(t)$. Because $T_{\text{max}}$ is finite, we expect these quantities to remain finite for $N\to\io$ in the
 R\&R-UNSAT phase, and to vanish in the R\&R-SAT phase (unless one is extremely close to the transition, in which case a larger $T_{\text{max}}$ would be needed). 
 We consider the minimal value because in this way one does not have to worry about the initial transient (e.g., if one wanted to take the average instead), 
 and the minimum has smaller fluctuations than the instantaneous value at $T_{\text{max}}$.
 
 In Fig.~\ref{fig:FSS1}  we report the average of $e_{\text{min}}$, $h_{\text{min}}$ and $z_{\text{min}}$ over samples in a finite size scaling plot. The scaling variable on the 
 horizontal axis
 is, as in a standard jamming transition, $\sqrt{N}(\sigma- \sigma_{\rm RR})$~\cite{baity2017emergent}. In the R\&R-UNSAT phase, we expect 
 $e_{\text{min}} \propto (\sigma-\sigma_{\rm RR})^2$,  $h_{\text{min}} \propto (\sigma-\sigma_{\rm RR})$,  $z_{\text{min}} \propto (\sigma-\sigma_{\rm RR})^0$, which
 fixes the scaling of the vertical axis. All data can be collapsed by a single free parameter, i.e. the value of $\sigma_{\rm RR}$, with deviations being
 observed only for the smallest size, $N=32$.
By this finite size scaling analysis, we can precisely estimate $\sigma_{\rm RR}=-0.228$ for $m=0$ and $\sigma_{\rm RR}=-0.175$ for $m=0.577$.
For other values of $m$, we estimated $\sigma_{\rm RR}(m)$ by collapsing data for $N=64$ and $N=128$ only, and we report the result in Fig.~\ref{fig:sigma_m}.

\subsection{Interpretation of the algorithmic transition}

\begin{figure}
\centering
\includegraphics[scale=0.65]{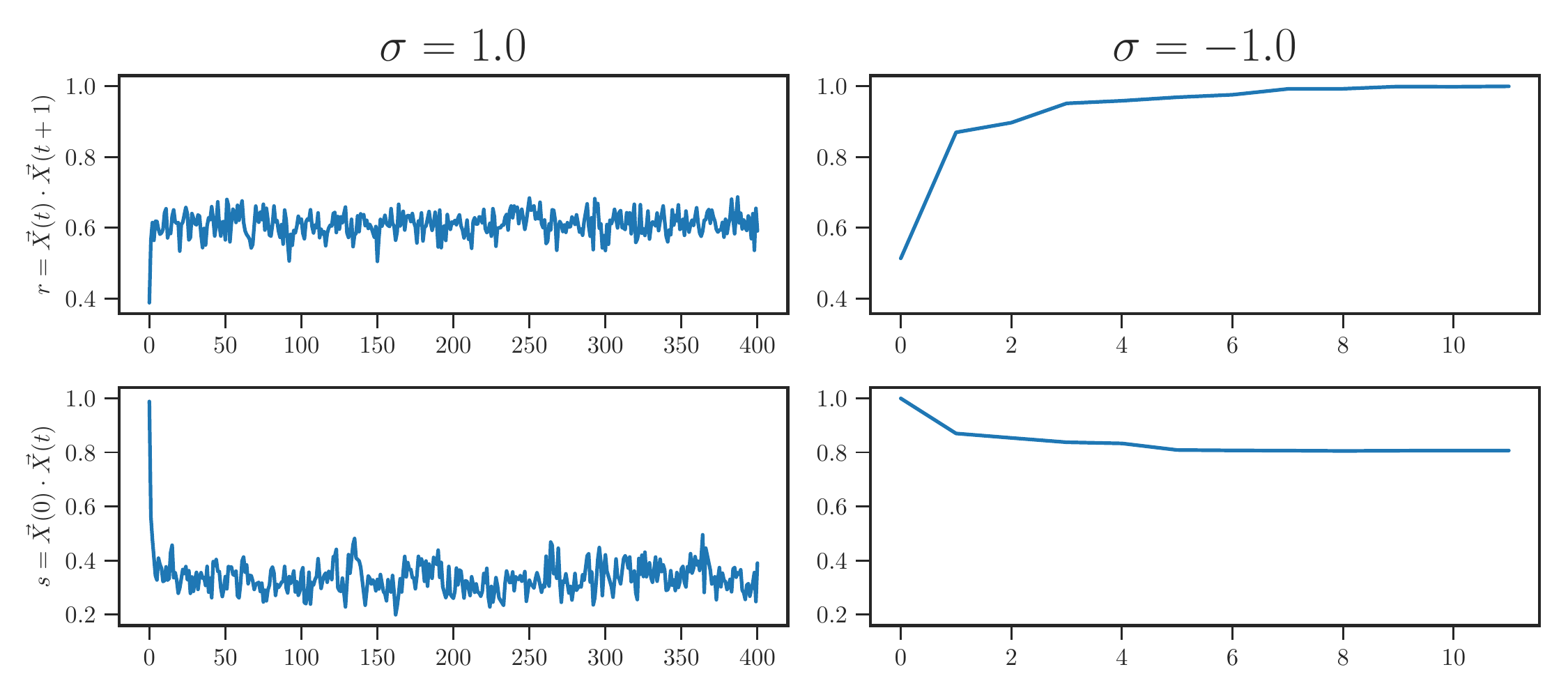}
\caption{
Overlaps $r(t)$ and $s(t)$ for $\a=20$, $m=0.577$, $N=200$, and two values of 
$\sigma = 1.0$ and $\sigma=-1.0$ in the R\&R-UNSAT and R\&R-SAT phases respectively.
}
\label{fig:overlaps_planting}
\end{figure}

We thus find that a sharp algorithmic transition at $\sigma_{\rm RR}(m)$ separates a R\&R-SAT phase from a R\&R-UNSAT phase for $N\to\io$.
Our interpretation of this transition is the following. If the value of $\vec X^{\rm opt}(t)$ does not change too much from one R\&R step to the next,
then R\&R becomes essentially equivalent to standard planting and quickly converges to a zero-energy state. 
In order to confirm this picture, we compute the following overlaps: 
\begin{align}
r(t) &:= \frac{1}{N} \vec{X}^{\text{opt}}(t+1) \cdot \vec{X}^{\text{opt}}(t) \ , \\
s(t) &:= \frac{1}{N} \vec{X}^{\text{opt}}(t) \cdot \vec{X}^{\text{opt}}(0) \ .
\label{overlap_planting}
\end{align}
$r(t)$ is the overlap between the optimum $\vec{X}^{\text{opt}}$ at two consecutive time steps of the R\&R dynamics and $s(t)$ is the overlap between the $\vec{X}^{\text{opt}}(t)$ and the initial optimal solution $\vec{X}^{\text{opt}}(0)$, corresponding to the initial choice of the disorder $\vec{\xi}_\mu$.
The numerical 
results for $r(t)$ reported in Fig.~\ref{fig:overlaps_planting} indeed confirm that in the R\&R-SAT phase $r(t)\to 1$,
indicating that $\vec{X}^{\text{opt}}(t)$ stabilizes to a constant value.
On the contrary, in the R\&R-UNSAT during the R\&R dynamics, at each step we are sampling significantly different realizations of the $\vec{\xi}_\mu$ and therefore new optimal solutions $\vec{X}^{\text{opt}}$. The dynamical process never converges and $\vec{X}^{\text{opt}}$ keeps evolving randomly on the sphere $\mathcal{S}_N$, with
$r(t) < 1$.

There are at least two mechanisms that can stabilize $\vec{X}^{\text{opt}}$ and thus ensure that self-planting is efficient. First of all, if $m$ is large, the space
of allowed $\vec X$ is very small, hence $\vec X$ cannot change too much from one step to the next. In the limiting case $m=1$, $\vec X \equiv (1,\cdots,1)$ is constant, and R\&R is equivalent
to planting. We expect that this mechanism dominates for $m$ close enough to one, and indeed $\sigma_{\rm RR}(m)\to \io$ when $m\to 1$, i.e., R\&R (like planting) 
becomes efficient
at all values of $\sigma$.

Another mechanism that can stabilize $\vec{X}^{\text{opt}}$  is the roughness of the energy landscape. Indeed, if the Hamiltonian has a single energy minimum (as in
the thermodynamic RS-UNSAT phase), then we expect the location of this minimum to depend sensitively on the disorder. Conversely, if the energy landscape
is rough, we expect that a local energy minimum is present close enough to each configuration $\vec X$. Therefore, in the RSB phase we would expect R\&R to converge
to zero energy more easily. 

\begin{figure}
	\includegraphics[scale=0.6]{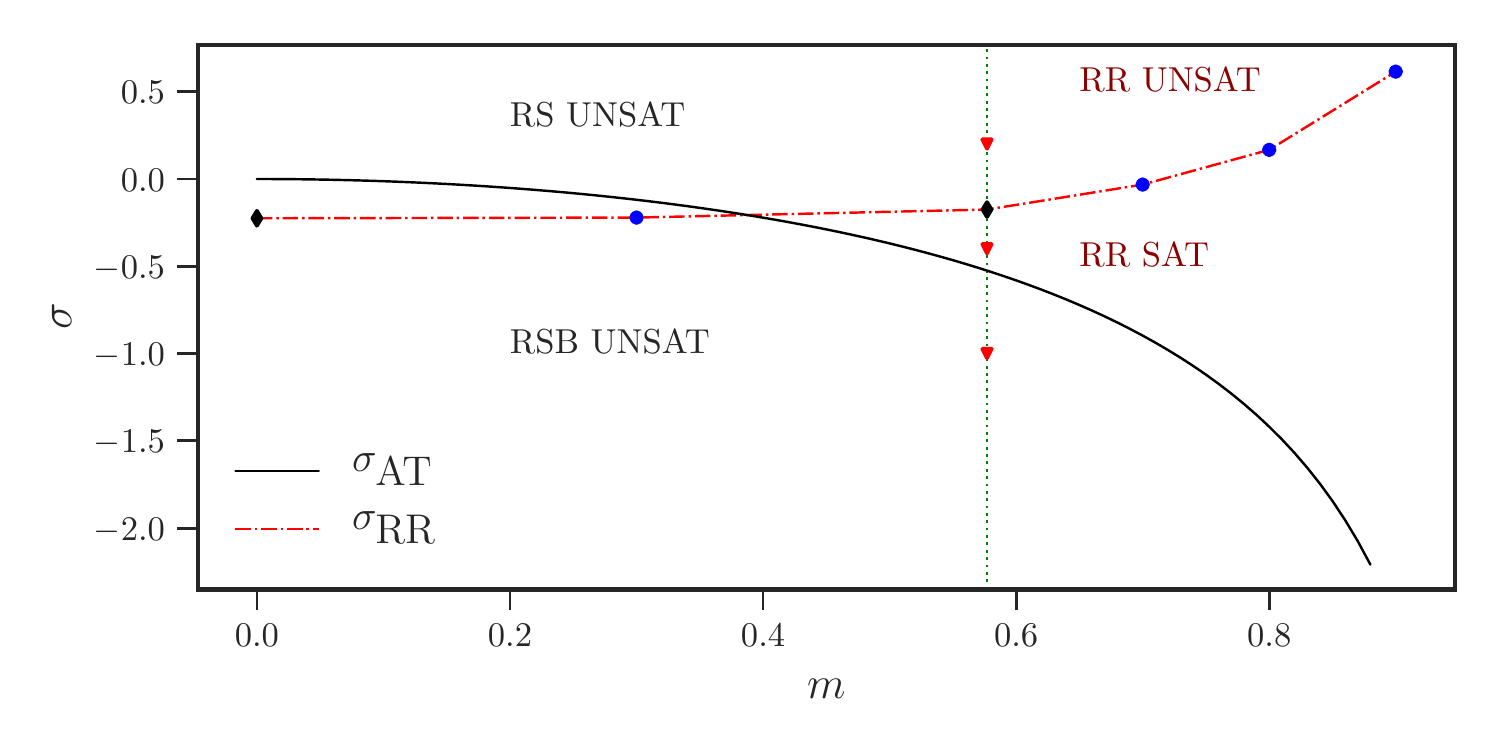}
\caption{
The full line is the de Almeida-Thouless transition $\sigma_{\rm AT}$, below which replica symmetry is broken in the UNSAT phase (RSB UNSAT),
obtained analytically for $N\to\io$.
Dots indicate the R\&R algorithmic transition $\sigma_{\rm RR}$, below which R\&R dynamics converges to a self-planted state (R\&R SAT),
obtained numerically from the finite size scaling (diamonds, with four values of $N$ as reported in Fig.~\ref{fig:FSS1}; circles, with two values of $N$, not shown),
for $\a=20$ and as a function of $m$. The vertical dashed line and triangles indicate the points that are further investigated in Fig.~\ref{fig:olaps_and_energies_temperature}.
}
\label{fig:sigma_m}
\end{figure}

In order to confirm this picture, in Fig.~\ref{fig:sigma_m} we compare the results for $\sigma_{\rm AT}(m)$ and $\sigma_{\rm RR}(m)$. At small $m$, we find that
the two values are indeed quite close. R\&R converges if started from the thermodynamic RSB phase, except very close to the de Almeida-Thouless transition. 
At larger $m$, R\&R becomes more and more efficient, and it converges even when initialized in the thermodynamic RS phase, presumably because the phase
space is restricted by the magnetization constraint.

Finally, one can ask whether the disorder $\vec{\xi}^\infty_\mu$ acquires a special structure in the self-planted state. A way to investigate this is to study the spectrum of the covariance matrix $C_{\mu \nu}= \vec{\xi}^\infty_\mu \vec{\xi}^\infty_\nu$. In the absence of any structure, we expect that the spectrum is of the Marcenko-Pastur type~\cite{franz2015universal}, whereas non-trivial structure would lead to some outliers (see e.g. \cite{Bouchaud2016a} for a recent review). We find that when $\sigma > 0$, the constraint $\vec X \cdot \vec \xi^\infty_\mu > \sigma \sqrt{N}$ is strong enough to induce a rank one correlation between the $\vec \xi^\infty_\mu$. When $\sigma < 0$, on the other hand, the most probable configurations are such that 
$\vec X \cdot \vec \xi^\infty_\mu \approx 0$ and no correlation is generated. In other words, we have not detected any non trivial correlation structure beyond that imposed by geometrical constraints. 

\subsection{Landscape exploration}

In an earlier section, we discussed the similarity between the phenomenology of the WalkSAT algorithm~\cite{Semerjian2003} and the R\&R dynamics. One significant way in which the current dynamics differs from the WalkSAT algorithm is in the evolution of the landscape itself. During R\&R dynamics, we observe that the energy landscape itself evolves and in what follows we discuss the structure that emerges from the dynamics.
To do so, we use the following procedure. At a given time $t$, the R\&R dynamics produces a pair $\{\vec{\xi}_\mu(t), \vec{X}^{\rm opt}(t)\}$.
The disorder $\vec{\xi}_\mu(t)$ encodes the energy landscape as defined by the Hamiltonian in Eq.~\eqref{define_hamiltonian}, while
$\vec{X}^{\rm opt}(t)$ is a {\it local} minimum of the Hamiltonian, selected by the dynamical history.
In order to characterize the energy landscape, for a fixed $\vec{\xi}_\mu(t)$,
we construct a series of random configurations $\vec{X}_{\text{new}}$ on the sphere $|\vec X|^2=N$, with constant magnetization $m=\sum_i X_i/N$,
and with given overlap $q_{\text{ini}}$ with $\vec{X}^{\rm opt}(t)$, 
i.e., such that 
\begin{equation}
q_{\text{ini}} = \frac1N \vec{X}^{\rm opt}(t) \cdot {\vec{X}_{\text{new}}}  \ .
\end{equation} 
Such new configurations are then evolved under standard gradient-descent (again, with fixed disorder). 
We then measure the overlap of each final configuration $\vec{X}_{\text{fin}}$ 
with the reference state,
\begin{equation}
q_{\text{fin}} = \frac1N \vec{X}^{\rm opt}(t) \cdot {\vec{X}_{\text{fin}}} \ .
\end{equation}
A scatter plot of $q_{\text{fin}}$ versus $q_{\text{in}}$ (called ``overlap plot'' in the following)
gives an idea of the structure of minima in the landscape.

For a simple RS landscape, there is a unique minimum, and all configurations fall in the unique minimum leading to
$q_{\text{fin}}=1$, irrespective of the initial distance $q_{\text{in}}$. The overlap plot is therefore a single horizontal line. If, on the contrary, the landscape is rough (as, for example, in a  fullRSB phase), then
the gradient descent procedure can terminate in one of many possible nearby minima. 
In this case, we expect that far away initial configurations, with $q_{\text{in}} \ll 1$, will fall in far away minima, $q_{\text{fin}}<1$, and we expect a monotonically
increasing overlap plot.

\begin{figure}[t]
\centering
\includegraphics[width=.7\textwidth]{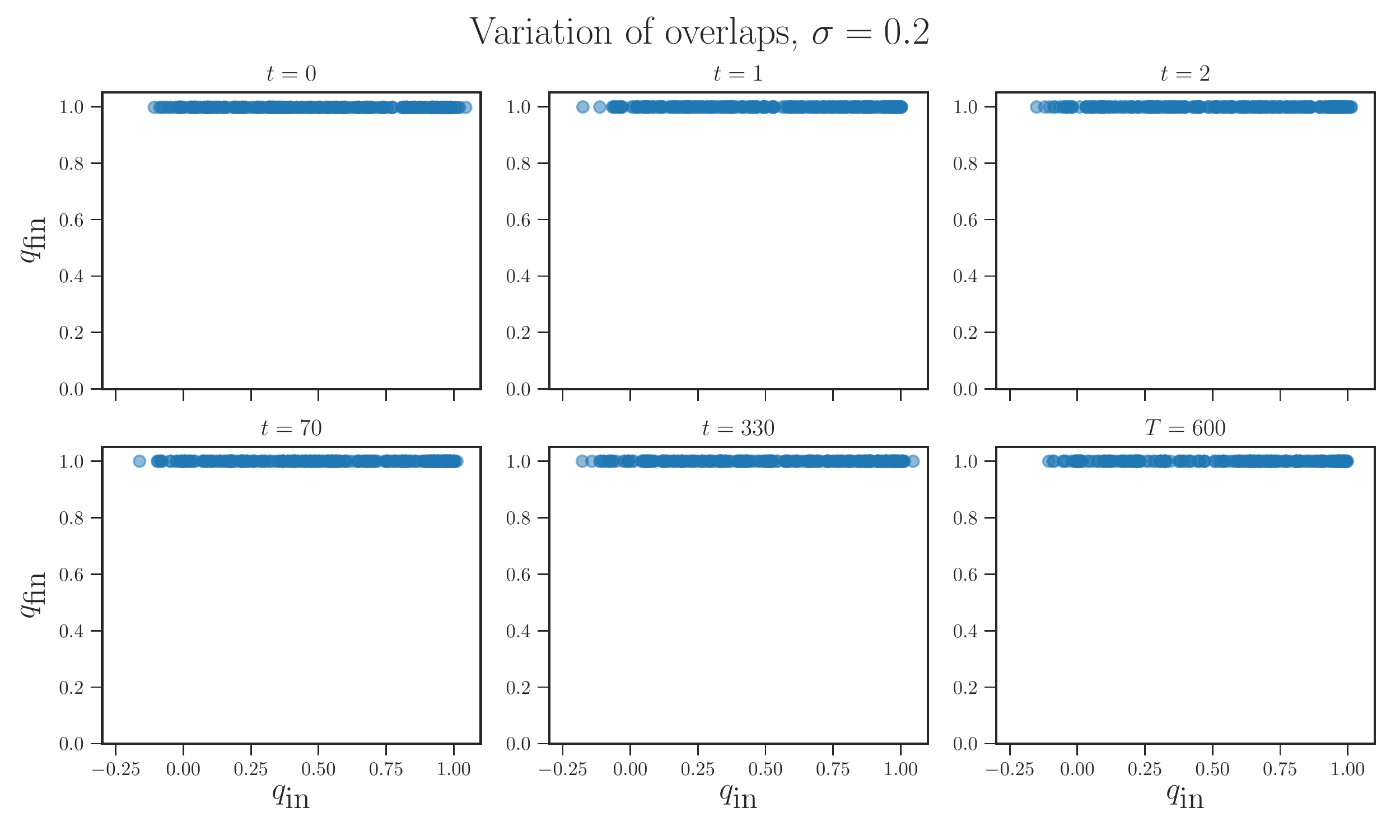}
\includegraphics[width=.7\textwidth]{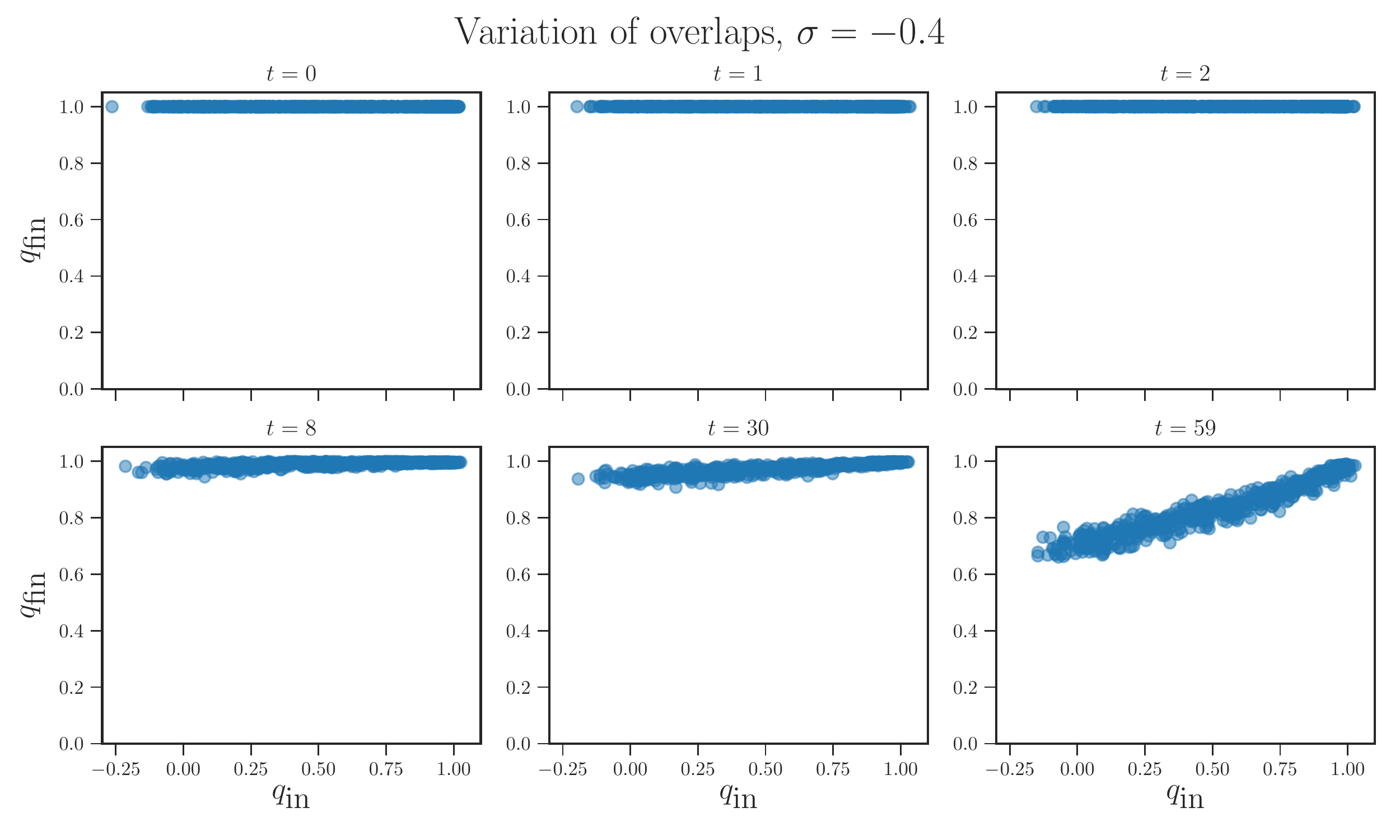}
\includegraphics[width=.7\textwidth]{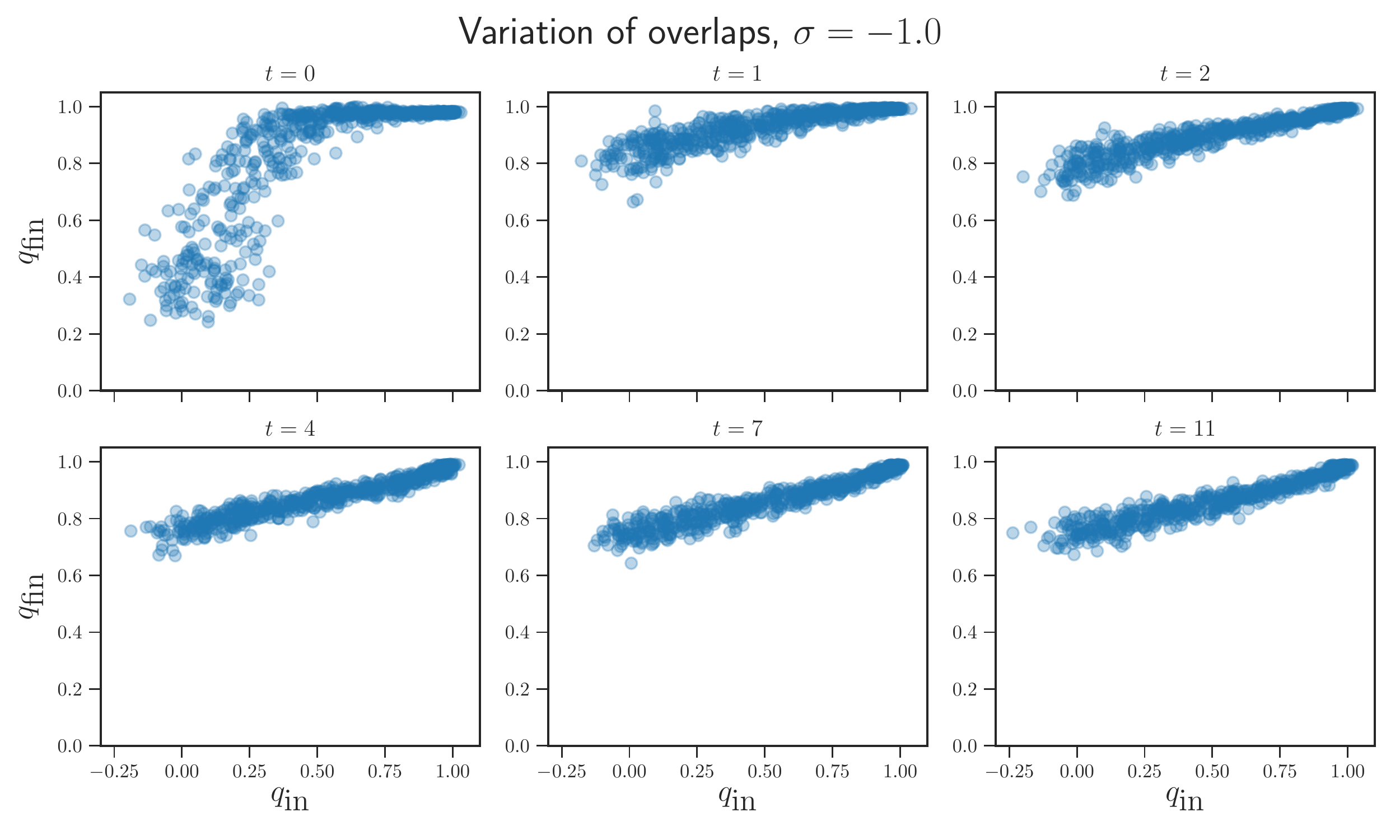}
\caption{Variation of the final overlaps $q_{\text{fin}}$ as a function of the initial overlap $q_{\text{in}}$, for $\a=20$, $N=200$, $m=0.577$. Overlap plots are shown for three distinct values of $\sigma$ corresponding to different R\&R and thermodynamic phases, as shown in Fig ~\ref{fig:sigma_m}. $\sigma=0.2$ is in the R\&R-UNSAT (thermodynamically RS) phase whereas $\sigma=-1.0$ is in the R\&R-SAT (thermodynamically RSB) phase. $\sigma=-0.4$ is an intermediate case: it is in the R\&R-SAT, but thermodynamically RS, phase. }
\label{fig:olaps_and_energies_temperature}
\end{figure}

In Fig.~\ref{fig:olaps_and_energies_temperature}, we show numerical results for the overlap plot, at selected values of $\sigma$ and $t$ in the different phases,
along the vertical dashed line of Fig.~\ref{fig:sigma_m} (red triangles):
\begin{itemize}
\item For $\sigma=0.2$, the initial state ($t=0$) is in a thermodynamic RS UNSAT phase. Indeed, we observe that the overlap plot is flat with $q_{\text{fin}}=1$,
indicating a trivial landscape with a single energy minimum. Under R\&R dynamics, we know that the energy remains finite and self-planting is not achieved.
We expect, as discussed above,
that this is due to the fact that the landscape remains simple, and its unique minimum changes strongly at each R\&R step, as indicated by $r(t)<1$ (Fig.~\ref{fig:overlaps_planting}).
Consistently, we observe that the overlap plot is unchanged at positive times $t>0$.

\item For $\sigma=-0.4$, the initial state is still thermodynamically RS and UNSAT, but now R\&R achieves self-planting, and  asymptotically the energy goes to zero
and $r(t)$ to one. In this case, we observe that the overlap plot is initially flat, but develops structure as time increases, indicating that the self-planting process
is accompanied by the development of a rough landscape, at least in the vicinity of the self-planted state.


\item Finally, for $\sigma=-1.0$ the initial state is thermodynamically RSB, and the initial overlap plot at $t=0$ is indeed monotonically increasing, indicating a rough landscape. For subsequent times, we observe that the landscape remains rough. However, for configurations with $q_{\text{in}} \sim 0$, final configurations are closer to the self-planted one. This indicates the presence of many low-energy minima, close to the self-planted one. 

\end{itemize}
Note that similar results in the planted case have been found in~\cite{Ros2018a}.

\section{Conclusions}

In this paper we have discussed the existence of a new sharp algorithmic transition (in the thermodynamic limit) for the R\&R dynamics applied to the perceptron model,
 which separates a R\&R-UNSAT phase at large $\sigma$ (in which the system is unable to produce a SAT, or self-planted configuration on finite time-scales) 
 from a R\&R-SAT phase at low $\sigma$ (in which a zero-energy configuration is found in a finite time). 
 Numerical evidence of such a transition is convincingly obtained by a finite-size-scaling analysis of the numerical data of finite-size samples. 
 We have shown that the transition to the SAT phase is generally associated to the fact that the optimal configuration does not evolve during the R\&R dynamics,
 so that it reduces to simple planting. This can be driven by (at least) two different mechanisms: 
  in the limit of large $m$ the space of allowed configurations 
 is very small, and $\vec X^{\rm opt}$ cannot change for trivial geometrical reasons. 
 Conversely, at small $m$ the satisfiability of the R\&R dynamics can be attributed to an emerging roughness of the energy landscape, 
 which produces a huge number of local minima and quasi-optimal configurations close to any initial condition, so that even if the disorder changes
  significanlty in a single R\&R step, there is always a new minimum close to the old one.



Self-planting is thus accompanied by a modification of the energy landscape. Exact analytical treatment of the phenomenon is difficult since the R\&R scheme couples the gradient
descent (or zero-temperature) dynamics of $\vec{X}$ under fixed disorder with an evolution of the disorder itself. Approximation schemes may perhaps be constructed to capture the phenomenon: one could for example only replace a small, random fraction of the violated constraints at each time step and set up a continuous-time approximation for the joint distribution of the $\vec{\xi}_\mu$ and $\vec{X}$. 

As mentioned in the introduction, it would be very interesting to see whether the self-planting phenomenon also occurs in other models characterized by a large connectivity of the constraints, such as the Hopfield model~\cite{Hopfield1982} or polydisperse soft spheres in high dimensions~\cite{charbonneau2017glass}. An equivalent to R\&R dynamics in the latter case would then be the removal and replacement of overlapping spheres after performing a gradient descent starting from a random initial condition. A particularly interesting setting to apply the R\&R construction is that of a classification problem. One could imagine a situation in which a large dataset of training examples $(\vec \Xi_\m, y_\m)$ is available, and a given class of classifiers parametrized by a vector $\vec X$ is used. The R\&R procedure would attempt, in this case, to find a subset of the whole dataset that is perfectly classifiable, through a co-evolution of the subset and
of the classifier itself. It would be interesting to check whether the resulting subset of examples and the resulting classifier have special and perhaps interesting properties. We leave this application for future work.

Finally, the interpretation of the perceptron as a market model and the consequences of the results presented here in terms of such macro-economic framework, in particular the existence of many solutions to the price setting problem, will be discussed in a forthcoming paper~\cite{Sharmaa}.

\acknowledgments

MT is a member of the Institut Universitaire de France. We thank J.~Kurchan and J.-P.~Nadal for interesting comments on this work, and we thank G.~Semerjian for suggesting
the use of the minimal value in the finite size scaling analysis of Fig.~\ref{fig:FSS1}.

\bibliographystyle{apsrev4-1}
\bibliography{paper_planting.bib}

\end{document}